\documentclass[twocolumn,notitlepage,nofootinbeib, aps, prl]{revtex4-2}
\usepackage{amsmath,amssymb}
\usepackage{ulem,cancel}
\usepackage{tempora}
\usepackage{newtxmath}
\usepackage[x11names]{xcolor}
\usepackage{latexsym,amsmath,amssymb,amsbsy,graphicx, float}
\usepackage[unicode,colorlinks,citecolor=Blue3,linkcolor=Blue3,bookmarks=true]{hyperref}

\newcommand\bra[2][]{#1\langle {#2} #1\rvert}
\newcommand\ket[2][]{#1\lvert {#2} #1\rangle}
\newcommand{\braket}[2]{ \langle #1 | #2 \rangle}

\begin{document}

	\title{
		Protocol for preparing Schr\"odinger-cat states via spontaneous parametric down-conversion and photon number measurement
	}

	\author{V.\,L.\,Gorshenin}
	\email{valentine.gorshenin@yandex.ru}
	\affiliation{Russian Quantum Center, Skolkovo 121205, Russia}
	\affiliation{Moscow Institute of Physics and Technology, 141700 Dolgoprudny, Russia}

	\begin{abstract}

		Schr\"odinger-cat (SC) states are an important resource for continuous-variable quantum computing and quantum metrology. In our previous work \cite{Gorshenin_JOSA_B_42_2_2025}, we proposed a probabilistic protocol for generating bright squeezed SC states via degenerate spontaneous parametric down-conversion (SPDC) with pump depletion, followed by projective measurement of the pump mode. In the present work, we formulate a general theoretical description of SPDC with pump depletion, introduce an efficient numerical method for computing its dynamics, and develop a practical version of the protocol proposed in \cite{Gorshenin_JOSA_B_42_2_2025}.

	\end{abstract}

	\maketitle

	\paragraph{Introduction.}

Schr\"odinger-cat (SC) states \cite{Dodonov_Phys_72_2_1974} are among the most paradigmatic non-Gaussian states, as reflected in the negativity of their Wigner function \cite{Schleich2001}. Their non-Gaussian character makes them especially appealing for quantum-information processing, because it guarantees the existence of states orthogonal to a given SC state, thereby enabling discrimination and encoding protocols \cite{Schleich2001, HelstromBook}.

SC states have long been recognized as a valuable resource for continuous-variable quantum computing \cite{T_C_Ralph_PRA_68_4_2003, T_C_Ralph_PhysRevLett_95_10_2005, G_Y_Xiang_NatPhotonics_4_5_2010, B_P_Lanyon_NatPhys_5_2_2009, A_P_Lund_PRL_100_3_030503_2008, A_I_Lvovsky_Arxiv_2006_16985_2020, M_Walschaers_PRX_Quant_2_3_2021, Gravina_PRXQuantum_4_020337_2023, Goto_SciRep_6_21686_2016}. They have also been explored in quantum metrology \cite{Shukla_OptQEl_55_460_2023, Shukla_PhOpen_18_100200_2024, Singh_PhOpen_18_100198_2024, V_L_Gorhsenin_21_6_2024, V_L_Gorshenin_Arxiv_2405_07049} and in macroscopic tests of quantum physics \cite{W_Marshall_PRL_91_14_2003, O_Romero-Isart_NewJofPhys_12_3_2010, F_Khalili_PRL_105_7_2010}.

An SC state is, by definition, a coherent superposition of two coherent states. In the classical limit, this structure is associated with bistability, namely the coexistence of two stable configurations. A natural setting for such physics is a cavity containing a nonlinear medium, whose dynamics is often described by two modes: the fundamental (signal) mode and the second-harmonic (pump) mode. The nonlinear interaction between these modes is commonly modeled as degenerate spontaneous parametric down-conversion (SPDC) \cite{Armstrong_PhysRev_127_6_1962}. Within the rotating-wave approximation, the corresponding interaction Hamiltonian reads
	\begin{equation} \label{SHG-hamiltonian}
		\hat{H}= \hbar g
		\left(\hat{a}^2 \hat{b}^\dagger + \hat{a}^{\dagger 2}\hat{b}\right)
	\end{equation}
	where \(\hat{a}\) and \(\hat{b}\) are the annihilation operators for the signal (first harmonic) and pump (second harmonic) modes, respectively, and \(g\) is the interaction coefficient. Throughout this Letter, we assume \(\hbar = 1\).

A full quantum treatment of the two-mode SPDC dynamics is nontrivial. Early progress came from quantum-trajectory simulations in the positive-P quasiprobability representation \cite{P_D_Drummond_JPhysA_13_7_1980, P_D_Drummond_Optica_Acta_27_3_1980}. An important insight was that, under suitable conditions, the pump mode---including coherent driving, optical loss, and parametric coupling to the first harmonic---can be adiabatically eliminated. The resulting effective description of the first-harmonic mode combines squeezing with two-photon dissipation and admits an SC state as a steady state \cite{L_Gilles_PRA_48_2_1993, L_Gilles_PRA_49_4_1994, E_E_Hach_III_PRA_49_1_1994}. This picture was later supported by numerical studies \cite{Nikitin_QOpt_JEOS_B_3_2_1991, Tanas_QOpt_JEOS_B_3_4__1991, Alvarez_JPA_28_20_1995}.

In Ref.~\cite{Gorshenin_JOSA_B_42_2_2025}, we showed that an SC state can be prepared by combining a SPDC interaction with a heralding measurement that projects the pump mode onto the vacuum state. Starting from a coherent state in the pump mode and vacuum in the signal mode, one can choose the interaction time such that conditioning on the outcome of zero pump photons yields an output state with fidelity approaching unity with the target SC state. Reliable heralding on a zero-photon outcome, however, may require additional techniques, such as the use of a reference pulse \cite{Nunn_PRA_104_3_2021}. Conditioning on nonzero photon-number outcomes is, by contrast, generally far more accessible experimentally. Moreover, Ref.~\cite{Gorshenin_JOSA_B_42_2_2025} did not provide a detailed physical explanation of the protocol. In the present Letter, we extend the protocol to allow conditioning on any small selected photon number, elucidate its underlying mechanism, and discuss prospects for experimental implementation.

  \paragraph{The Nikitin-Masalov representation.}

	The Nikitin--Masalov representation \cite{Nikitin_QOpt_JEOS_B_3_2_1991} relies on the conservation law that follows from the commutation of the Hamiltonian \eqref{SHG-hamiltonian} with the total-energy operator:
	\begin{equation}\label{eq:total-energy-operator}
		\hat{N} = \hat{a}^\dagger \hat{a} + 2 \hat{b}^\dagger \hat{b}
	\end{equation}
	
	Accordingly, the dynamics generated by \eqref{SHG-hamiltonian} decomposes into invariant subspaces with fixed \(N = n_{\rm s} + 2 n_{\rm p}\), so that the wave function may be expanded independently in each such subspace:
	\begin{equation}\label{eq:NM-repres-defenition}
		\ket{\psi} = \sum_{N=0}^{\infty} \sum_{k=0}^{k = [N/2]} \psi_{N,k} \ket{N-2k}_{\rm s} \otimes \ket{k}_{\rm p} \,,
	\end{equation}
	where the subscripts \({\rm s}\) and \({\rm p}\) refer to the signal and pump modes, respectively. In this representation, the Hamiltonian reduces, in each fixed-\(N\) subspace, to a tridiagonal matrix (see Eq.\,\eqref{tridialgonal-hamiltonian}). Its eigenvalues and eigenstates are denoted by \(\lambda^{N, j}\) and \(\ket{\chi^{N,j}}\), respectively. Additional details are given in \cite{supplement}.

	In the Schr\"odinger picture, we represent the evolving state by expanding it in the eigenbasis of the Hamiltonian:
	\begin{equation}
		\ket{\psi(t)}
		= \sum_{N=0}^\infty
		\sum_{j=0}^{j=[N/2]} e^{-i \tau \lambda^{N,j}}
		\ket{\chi^{N,j}}\braket{\chi^{N,j}}{\psi(t=0)} \,,
	\end{equation}
	where
  \begin{equation}
    \tau = gt
  \end{equation}
  is the dimensionless time. The initial two-mode state is:
  \begin{equation}
  	\ket{\psi(t=0)} = \ket{0}_{\rm s} \otimes \ket{\beta}_{\rm p}\,,
  \end{equation} 
  where \(\beta\) -- real and positive amplitude of initial coherent state in the pump mode. This state can be written as follows:
	\begin{equation}
		\ket{\psi(t=0)} =
		e^{-\beta^2 / 2}\sum_{N{\rm-even}, N\ge0}^{\infty} \frac{\beta^{N/2}}{\sqrt{(N/2)!}}\ket{0}_{\rm s} \otimes \ket{N/2}_{\rm p}
	\end{equation}
	\paragraph{Evolution of the system.}

  	In the Nikitin--Masalov representation, it suffices to analyze the dynamics starting from a two-mode Fock state such as \(\ket{0}_{\rm s} \otimes \ket{n}_{\rm p}\).

	The expansion of the initial state \(\ket{0}_{\rm s} \otimes \ket{n}_{\rm p}\) is dominated by a relatively small number of Hamiltonian eigenvectors with eigenvalues clustered symmetrically around zero. We therefore retain only a truncated set of eigenvalues near the center of the spectrum. Denoting by \(\rho\) the number of retained eigenvalues on either side of zero, we keep \(\lambda^{N, j}\) with \(j \in \big[[N/4]-\rho, [N/4]+\rho\big]\), where \(N = 2n\). We have verified that choosing \(\rho=25\)  is sufficient for our numerical calculations for \(n \leq 10^4\). A justification of this near-zero truncation is provided in  Supplemental Material \cite{supplement}.

	The asymptotic behavior of the eigenvalues of this Hamiltonian at large \(|\lambda|\) has been analyzed in Ref.~\cite{Alvarez_JPhysA_28_20_1995}; however, the structure of eigenvalues near zero was not considered there. In Ref.~\cite{Karassiov_PhysLettA_295_5_2002}, an approximation for the full spectrum was proposed. Unfortunately, for the eigenvalues near zero (except for \(\lambda = 0\)), that method yields a significant error. In \cite{supplement}, we present numerical approximations for the near-zero eigenvalues as functions of the even total photon number \(N\).

	At \(t=0\), all excitations reside in the second-harmonic (pump) mode, so the conserved quantity is \(N = 2n\). Expanding the initial state \(\ket{0}_{\rm s} \otimes \ket{n}_{\rm p}\) in the Hamiltonian eigenvectors, we consider the projection amplitudes \(\chi^{2n, j}_n = \bra{\chi^{2n, j}} \ket{0}_{\rm s}\ket{n}_{\rm p}\). The two-mode state \(\ket{\psi_{\rm 2\,mode}}\) then takes the form:

	\begin{equation}
		\ket{\psi_{\rm 2\,mode} (\tau)}
		=
		\sum_{j=[n/2]-\rho}^{j = [n/2]+\rho} \chi^{2n, j}_n \exp(-i \lambda^{2n,j} \tau) \ket{\chi^{2n, j}}\,,
	\end{equation}

  After conditioning on a projective measurement of the pump-mode photon number that yields the outcome \(m\), the two-mode state is projected onto the following state:
	\begin{equation}
			\ket{\psi_{\rm collapse}^{(m)} (\tau)}_{\rm s}
			\equiv
			\,_{\rm p} \langle m\ket{\psi_{\rm 2\,mode}}
			=
			A_{\rm n, m} \ket{2(n - m)}_{\rm s} \,,
	\end{equation}
  where
	\begin{equation}
		A_{\rm n, m} (\tau)
		\equiv
		\sum_{j=[n/2]-\rho}^{j = [n/2]+\rho} M^{n, j}_m(\tau)\,,
	\end{equation}
	are the transition amplitudes, and \(M^{n, j}_m(\tau)\) denotes the corresponding finite-sum contributions. Substituting the eigenstate components from Eq.\,\eqref{eq:eigenvector-components}, we obtain:
	\begin{equation} \label{eq:M_n_k_final}
		M^{n, j}_m(\tau) \equiv
		\chi^{2n, j}_n \chi^{2n, j}_m
		e^{-i \lambda^{2n,j} \tau}
	\end{equation}

	Because the dynamics is restricted to a fixed-\(N\) subspace, a pump-mode photon-number measurement yielding the outcome \(m\) projects the signal mode onto the Fock state \(\ket{2(n-m)}_{\rm s}\).

  \paragraph{Conditioning on $m=0$.} 
 
 In our previous work (Ref.~\cite{Gorshenin_JOSA_B_42_2_2025}), we introduced the optimal interaction time \(\tau_{\rm opt}\) as the interaction time for which the probability of measuring zero photons after the interaction, \(m = 0\), attains its first maximum. This notion extends naturally to small measured photon numbers \(m\). For small even \(m\), the corresponding dimensionless time is approximately given by the following expression (see Table 1 in \cite{supplement} for details):
  \begin{equation}
  	\tau_{\rm opt}^{(m)} \approx b^{(m)} / \beta^{d^{(m)}} \approx 1.2 / \beta^{0.8}
  \end{equation}

Following the interaction and the projective measurement, the transition amplitudes display a parity-selective structure: for even \(n\) one quadrature component (real or imaginary) vanishes, whereas for odd \(n\) the complementary quadrature vanishes. As a result, the postselected probability amplitudes are purely real for one parity and purely imaginary for the other. In addition, neighboring transition amplitudes acquire a relative phase shift of \(-\pi/4\) as previously observed in Ref.~\cite{Gorshenin_JOSA_B_42_2_2025}.

Applying the phase-space rotation \(\hat{R}(\phi) = \exp(-i\phi \hat{n})\) with \(\phi = - \pi/4\), makes all postselected probability amplitudes real and non-negative. We therefore define the corresponding phase-rotated signal-mode state as \(\ket{\psi_{\rm res}^{(m)}}\):
	\begin{equation}\label{eq:psi-res-def}
		\ket{\psi_{\rm res}^{(m)}} \equiv e^{i \frac{\pi}{4} \hat{n}} \Big|\psi_{\rm collapse}^{(m)} (\tau)\Big\rangle
	\end{equation}

For relatively large \(\beta\) (\(\beta \geq 5\)), the state \(\ket{\psi_{\rm res}^{(0)}}\) is well approximated by a squeezed Schr\"odinger-cat (SSC) state:
\begin{equation}\label{eq:cat-state-wavefunction-approx}
	\ket{\psi_{\rm SSC}^{(0)}}
	=
	K
	\hat{S}(r)
	\Big(\ket{\alpha} + \ket{-\alpha}\Big)\,,
\end{equation}
where \(K = (2(1 + e^{-2 \alpha^2}))^{-1/2}\) -- is the normalization factor, \(\alpha = \beta\) and \(r \approx - 0.35\) corresponding approximately to \(e^{-2r} \approx 2\). These parameter values were obtained numerically in Ref.~\cite{Gorshenin_JOSA_B_42_2_2025}.

\begin{figure}
		\centering
		\includegraphics[width=0.93\linewidth]{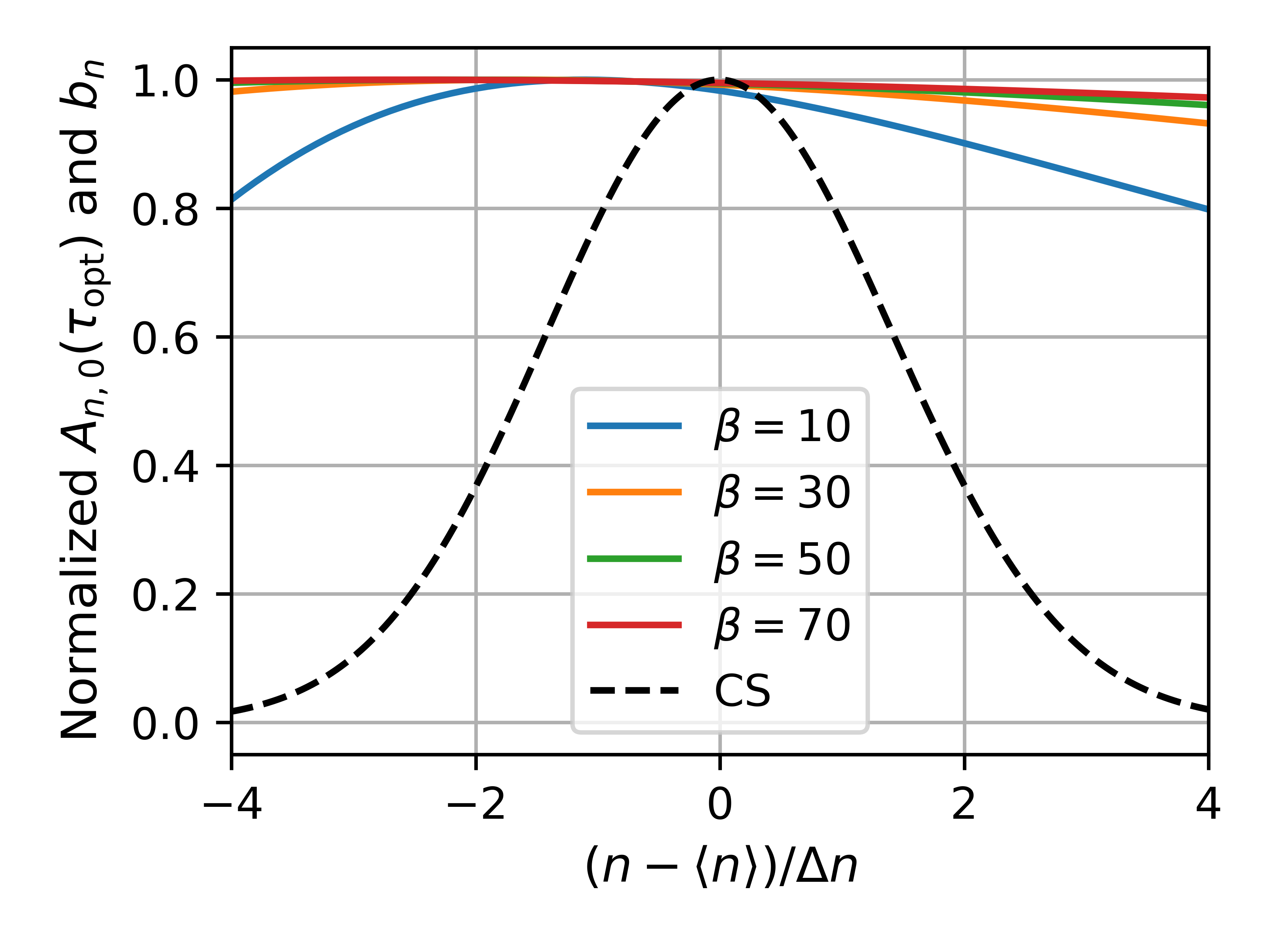}
		\caption{
			Transition amplitude \(A_{n, 0}(\tau)\) for the zero-photon pump outcome, evaluated at
			 \(\tau = \tau_{\rm opt}(\beta)\)), together with the corresponding coherent-state probability amplitudes \(b_n\) for \(\ket{\beta}\) (marked as CS - coherent state), for several values of \(\beta\).
					}
		\label{fig:conversion_and_coh_beta_array_tau_opt}
	\end{figure}

Our calculations further show that, for large initial coherent-state amplitude \(\beta\) the relevant transition amplitudes become nearly equal. In this regime, they act as conversion coefficients that map the initial wave function onto the prepared state:
	\begin{equation} \label{eq:prepared-state}
		\ket{\psi_{\rm res}^{(0)}}
		=
		e^{-\beta^2 / 2}\sum_{n=0}^{\infty} A_{n, 0} \frac{\beta^n}{\sqrt{n!}} \ket{2n}
	\end{equation}
Fig.\ref{fig:conversion_and_coh_beta_array_tau_opt} shows the transition amplitudes for Fock states conditioned on the projective measurement, evaluated at the optimal interaction time, together with the dominant part of the initial coherent-state wave function, whose coefficients are: \(b_n \equiv \exp(-\beta^2/2) \beta^n/\sqrt{n!}\). To compare different values of \(\beta\), we rescale the horizontal axis to the dimensionless variable \((n - \langle n \rangle) / \Delta n\), where \(\langle n \rangle = \beta^2\) and \(\Delta n =\beta\). For visual clarity, the transition amplitudes are normalized. Over the range in which the coherent-state weight is appreciable, the transition amplitudes remain nearly constant.

For constant  \(A_{n,0}\) the state \(\ket{\psi_{\rm res}^{(0)}}\) closely resembles the state \(\ket{\psi_{\rm SSC}^{(0)}}\) with \(e^{-2r} = 2\). This follows from three simple observations:

(i) Both \(\ket{\psi_{\rm res}^{(0)}}\) and \(\ket{\psi_{\rm SSC}^{(0)}}\) have non-zero pobability amplitudes only on even Fock states with arbitary non-zero \(A_{n,0}\) and small \(r\) (\(e^{|r|} \ll \alpha\)).

(ii) The two states have the same mean photon number. For \(\ket{\psi_{\rm res}^{(0)}}\), the initial mean photon number in the pump mode is \(\beta^2\). Under the mapping \(\ket{n}_{\rm p} \rightarrow \ket{2n}_{\rm s}\), this acquires an additional factor of 2, giving a mean photon number \(2 \beta^2\) in the signal mode. 
For \(\ket{\psi_{\rm SSC}^{(0)}}\), the mean photon number can be estimated as follows (see Eq.\,(36) in Ref.\,\cite{supplement}):
\begin{equation} \label{eq:n-avg-cat}
	\begin{aligned}
		\langle \hat{n} \rangle
		=
		\bra{\psi_{\rm SSC}^{(0)}} \hat{a}^\dagger \hat{a} \ket{\psi_{\rm SSC}^{(0)}}
		=
		e^{-2r} \alpha^2
	\end{aligned}
\end{equation}

\begin{figure}
	\centering
	\includegraphics[width=0.93\linewidth]{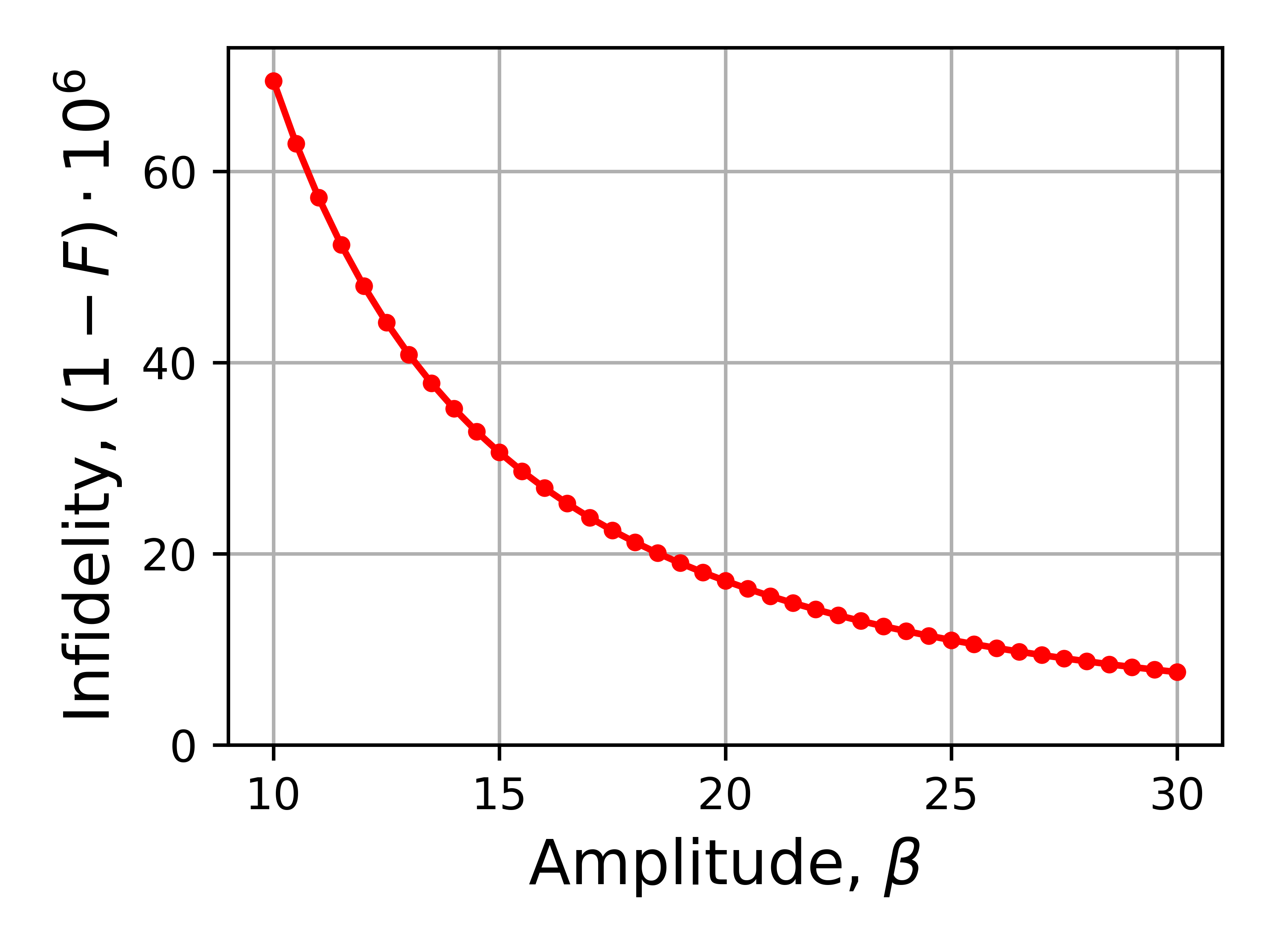}
	\caption{
		Infidelity, \(1 - F\) as a function of \(\beta\) for \(\ket{\psi_{\rm res}^{(0)}}\) with constant \(A_{n,0}\) (see Eq.\,\eqref{eq:prepared-state}) and \(\ket{\psi_{\rm SSC}^{(0)}} \) with \(e^{-2r} = 2\) and replacement \(\alpha \rightarrow \beta\) (see Eq.\,\eqref{eq:cat-state-wavefunction-approx}).
	}
	\label{fig:fidelity_output_and_sq_cat}
\end{figure}

(iii) 
Two states have the same photon-number variance \((\Delta n)^2\). For state \(\ket{\psi_{\rm res}^{(0)}}\), the remapping \(\ket{n}_{\rm p} \rightarrow \ket{2n}_{\rm s}\) increases the variance by a factor of 4 relative to the initial pump coherent state. Hence, \((\Delta n)^2 = 4 \beta^2\). For \(\ket{\psi_{\rm SSC}^{(0)}}\), the photon-number variance can be evaluated as follows (see Eq.\,(38) in Ref.\,\cite{supplement}):
\begin{equation}
	\begin{aligned}\label{eq:dispersion_n_approx}
		(\Delta n)^2 =
		\bra{\psi_{\rm SSC}^{(0)}} \hat{n}^2
		\ket{\psi_{\rm SSC}^{(0)}}
		-
		\big(\bra{\psi_{\rm SSC}^{(0)}}
		\hat{n}
		\ket{\psi_{\rm SSC}^{(0)}}\big)^2
		=
		e^{-4r} \alpha^2
	\end{aligned}
\end{equation}
Thus, conditions (ii) and (iii) independently yield the same squeezing parameter, \(e^{-2r} = 2\) and amplitude \(\alpha = \beta\), in agreement with Ref.~\cite{Gorshenin_JOSA_B_42_2_2025}. Fig. \ref{fig:fidelity_output_and_sq_cat} shows the fidelity deviation from unity(infidelity), \(1-F\), between \(\ket{\psi_{\rm res}^{(0)}}\) for constant \(A_{n,0}\) and \(\ket{\psi_{\rm SSC}^{(0)}}\) with \(e^{-2r} = 2\) upon setting \(\alpha \rightarrow \beta\) as a function of \(\beta\). 	

	\paragraph{Conditioning on $m>0$.}

Conditioning on a zero-photon detection event may be experimentally demanding, for example because photon number resolving detectors have nonzero dark-count rates. We therefore extend the analysis in this Letter to conditioning on nonzero photon-number outcomes in the pump mode after evolution under the Hamiltonian (see Eq.\,\eqref{SHG-hamiltonian}). Our goal is to generate the SSC state as quickly as possible during the nonlinear interaction, thereby minimizing the effect of losses. At the same time, measuring a small number of photons (up to 6) is experimentally feasible, whereas resolving larger photon numbers remains challenging.

\begin{figure}
		\centering
		\includegraphics[width=0.93\linewidth]{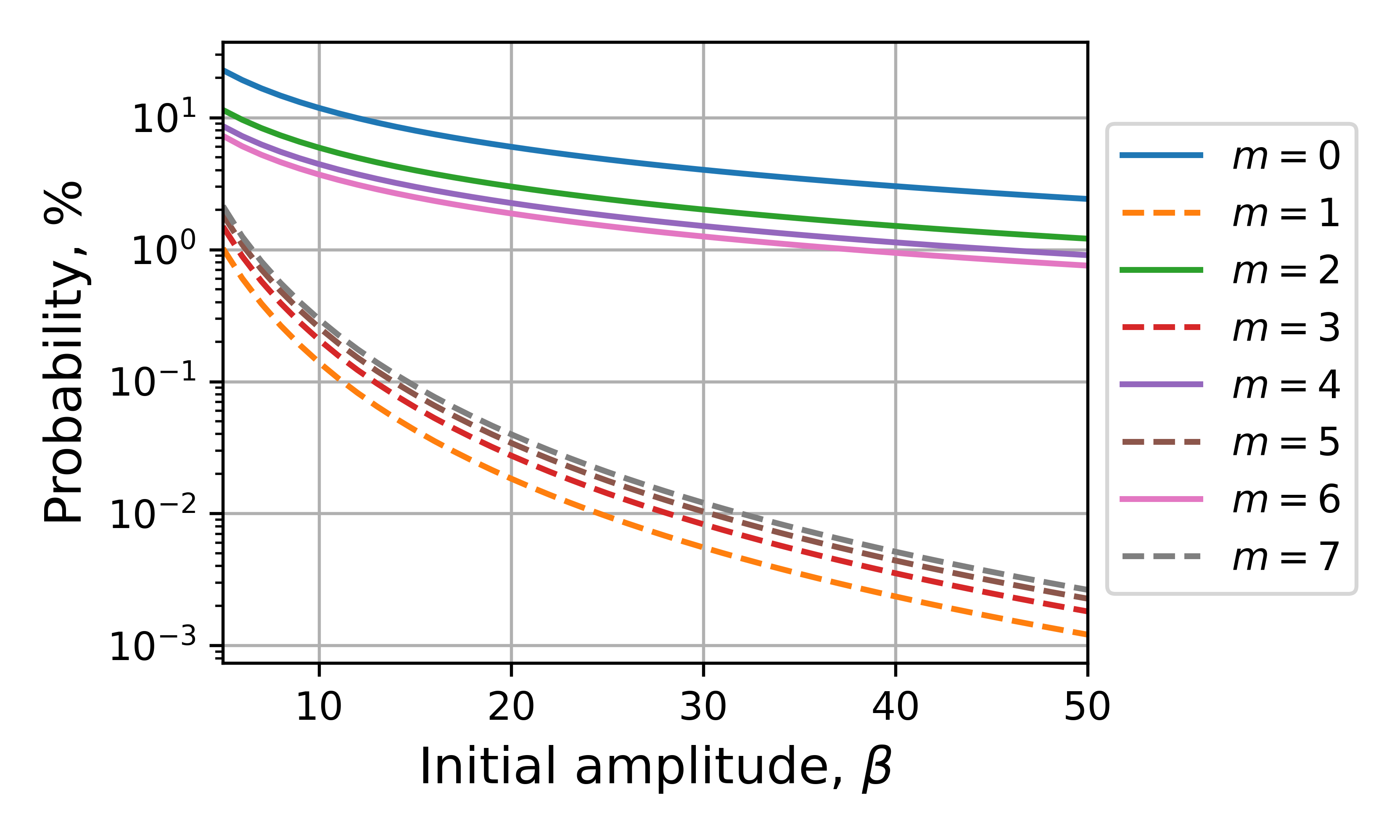}
		\caption{
			First maximum of the achievable probability of successfully measuring \(m\) photons in the pump mode as a function of the initial pump coherent-state amplitude \(\beta\).
					}
		\label{fig:prob-opt-time-e-n-o}
	\end{figure}

	The postselection probability for small photon numbers exhibits a pronounced even--odd asymmetry: even outcomes occur with substantially higher probability than odd ones. The corresponding maximal success probabilities are shown in Fig.~\ref{fig:prob-opt-time-e-n-o} for amplitude range \(5 \leq\beta \leq 50\). For this reason, we focus below on conditioning on small even photon-number outcomes only.

	Because the measurement may yield different photon numbers \(m\), we approximate the corresponding postselected state by the following SSC wave function:
	\begin{equation}\label{eq:cat-state-wavefunction-res-state-approx}
		\ket{\psi^{(m)}_{\rm SSC}}
		= K^{(m)}
		\hat{S}(r^{(m)})
		\big(\ket{\alpha^{(m)}} + \ket{-\alpha^{(m)}}\big)\,,
	\end{equation}
	where \(K^{(m)} = 2(1 + e^{-2 (\alpha^{(m)})^2})^{-1/2}\) -- normalization factor.
	
\begin{figure}
		\centering
		\includegraphics[width=0.93\linewidth]{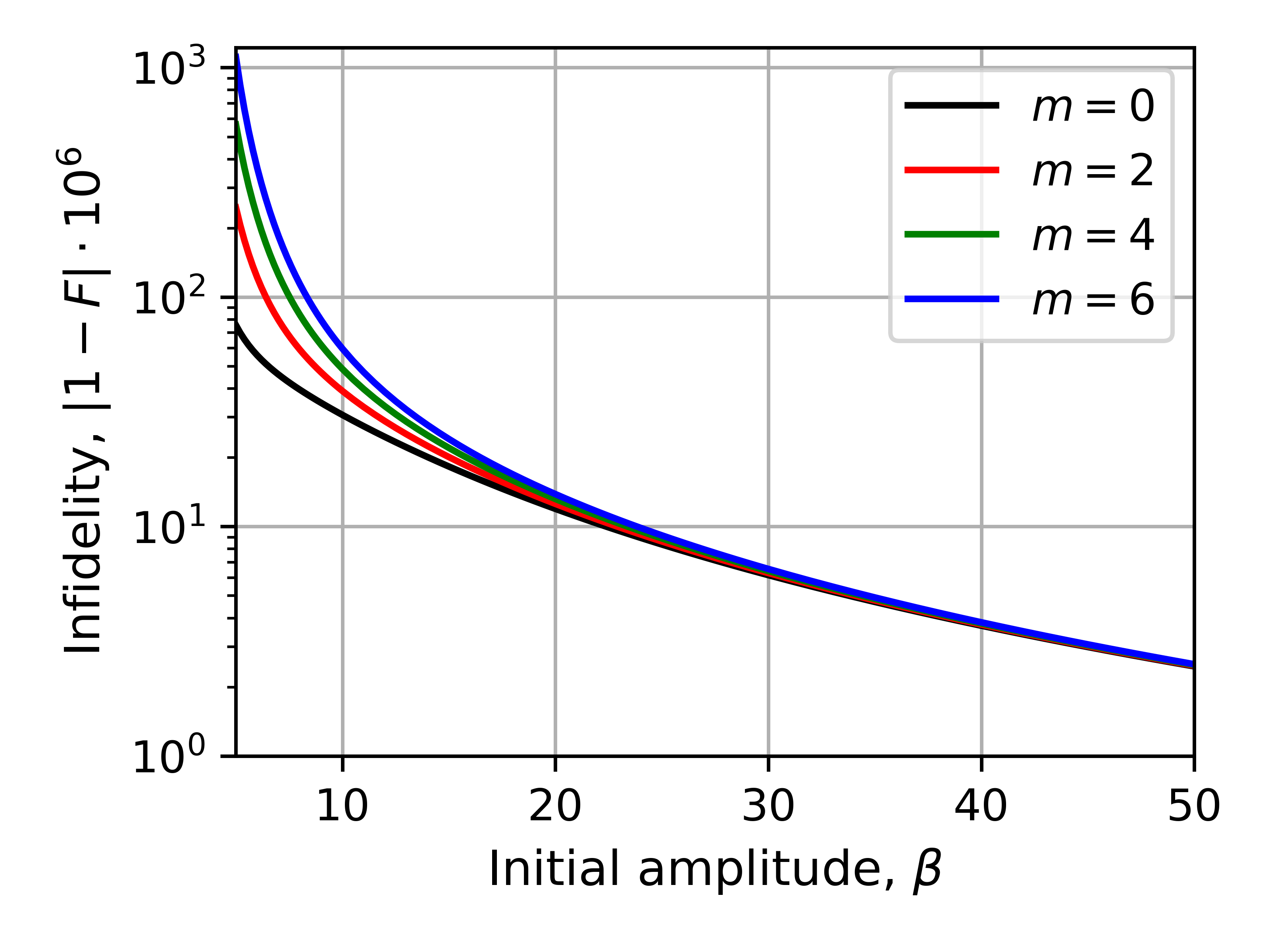}
		\caption{
			Fidelity deviation from unity (infidelity) of the postselected output state \(\ket{\psi_{\rm res}^{(m)}}\) with the corresponding squeezed cat state \(\ket{\psi_{\rm SSC}^{(m)}} \), conditioned on detecting a small pump photon number \(m\) after evolution for \(\tau_{\rm opt}(\beta)\), plotted versus the initial pump coherent amplitude \(\beta\).
		}
		\label{fig:fid-opt-time-e-n-o}
	\end{figure}

	We find that the proposed projection method can prepare an SC state with high fidelity (\(F>99.99\%\)) when small even photon numbers (up to 6 photons) are measured in the pump mode after an interaction time \(\tau_{\rm opt}\). Fig.~\ref{fig:fid-opt-time-e-n-o} shows the infidelity between \(\ket{\psi_{\rm res}^{(m)}}\) and \(\ket{\psi^{(m)}_{\rm SSC}}\) as a function of the initial coherent-state amplitude, for several even photon measurement outcomes \(m\) on pump mode. Across the entire parameter range considered (\(5\leq\beta\leq50\)), the fidelity remains close to unity.

	For a nonzero pump-photon outcome \(m\),  the amplitude \(\alpha^{(m)}\) of the resulting SSC state \(\ket{\psi^{(m)}_{\rm SSC}}\) is well approximated by:
	\begin{equation}\label{eq:connection-init-amplitude-and-ssc-amplitude}
		\alpha^{(m)}(\beta) \approx \sqrt{\beta^2 - m}
	\end{equation}

	The corresponding relation, together with numerical estimates of the squeezing parameter (\(- 0.3 \leq r^{(m)}\leq-0.4\)) as a function of \(\beta\), is given in \cite{supplement}.

  \paragraph{Discussion.}

The proposed protocol requires two key ingredients: degenerate SPDC interaction and photon-number-resolving detection of the pump mode. 
Below we discuss three representative platforms that offer these capabilities: nonlinear optical resonators, trapped-ion phonon modes in Paul traps, and superconducting Josephson-junction circuits coupled to microwave resonators.

In the optical domain, strong second-order nonlinearity (\(\chi^{(2)}\)) can be achieved in noncentrosymmetric materials such as lithium niobate and lithium tantalate. Considerable recent progress has been made in the fabrication of high-\(Q\) microresonators based on these materials \cite{Gao_NewJPhys_23_123027_2021, Chen_PhysRevAppl_16_064004_2021, Zhang_Optica_4_12_2017, He_PhotonicsResearch_13_5_1385_1389_2025}. In particular, lithium niobate microresonators with \(Q \sim 10^8\) have been demonstrated \cite{Gao_ChineseOpticsLetters_20_1_011902_2022}. For representative parameters, our estimate in \cite{supplement} yields a maximum dimensionless interaction time of \(\tau \lesssim 0.25\), which may already be sufficient for generating relatively bright and bright SSC states.

A second route is to engineer an effective degenerate SPDC interaction in centrosymmetric platforms, such as silicon nitride, by exploiting an underlying degenerate four-wave-mixing process together with a strong pump. In this setting, three modes are selected, and the pumped mode is treated classically, leading to an effective two-mode Hamiltonian of the form \eqref{SHG-hamiltonian}. This approach has enabled both degenerate and nondegenerate effective parametric interactions \cite{Chen_PhysRevAppl_16_064004_2021, Ramelow_RPL_122_153906_2019}. As estimated in \cite{supplement}, representative parameters allow \(\tau \lesssim 0.017\), indicating that bright SSC states may also be accessible in this setting.

The second key requirement for optical implementation is a PNR detector. For a native degenerate SPDC process with a 1550-nm signal, the pump wavelength lies near 780 nm, where transition-edge-sensor (TES) PNR detectors are a natural choice \cite{Lita_AIPConferenceProceedings_1185_1_2009, Fukuda_OptExpress_19_2_870_875_2011}. For an effective SPDC interaction, the pump mode is instead near 1550 nm, in which case either TES-based PNR detectors \cite{Li_JournalOfLowTemperaturePhysics_209_3_248_255_2022, Hattori_IEEETransactionsOnInstrumentationAndMeas_68_6_2253_2259_2019} or nanowire-based PNR detectors \cite{Ding_ACSPhotonics_12_4294913_2025} may be employed.

In trapped-ion phononic systems, degenerate parametric interactions can arise from nonlinear terms obtained by expanding the Coulomb interaction about the ions’ equilibrium positions \cite{Ding_PRL_119_15_2017}. Phonon-number measurement schemes have likewise been proposed and experimentally demonstrated \cite{Mallweger_PRL_131_22_2023}. For \(Yb^{171+}\) ions, the nonlinear interaction coefficient can reach \(g\approx 6.5\cdot 10^3\)  \cite{Ding_PRL_115_2017}, while in that experiment the signal-mode angular frequency was \(\omega_s \approx 4 \cdot 10^6\). The usable interaction time is ultimately limited by phonon relaxation, which we approximate here by the heating time of the mode. Using the worst-case heating rate of 25 quanta/s reported in Ref.~\cite{Spivey_IEEETransOnQuantEngineering_3_2021}, we obtain an effective relaxation time \(t_{\rm rel} \approx 20\) ms which corresponds \(\tau \lesssim 1.3\cdot10^2\). This estimate strongly supports the feasibility of SSC-state generation in trapped-ion phononic platforms.

In the microwave domain, degenerate parametric interactions can be implemented using the intrinsic nonlinearity of Josephson-junction circuits \cite{Majumdar_PRB_87_23_2013, Shen_H_PRA_90_2_2014, Cao_PRA_84_5_2011}. Photon-number statistics may be measured with the aid of an ancillary Josephson-junction qubit \cite{Johnson_NatPhys_6_9_2010, Curtis_PRA_103_2_2021}. As a concrete example, we consider the SSC-state generation scheme of Ref.~\cite{Z_Leghtas_Science_347_6224_2015}, which reports a nonlinear interaction coefficient \(g\approx7\cdot10^5\), a signal-mode angular frequency \(\omega_s \approx 5\cdot 10^{10}\) and a signal-mode relaxation time \(t_{\rm rel} \approx 20\) \(\mu\)s. If the pump mode relaxes on a comparable timescale, then the maximum dimensionless interaction time is \(\tau \lesssim 14\). This estimate strongly supports the feasibility of SSC-state generation in superconducting platform.

	\acknowledgments

	The  numerical evolution calculation development was supported by Rosatom in the framework of the Roadmap for Quantum computing (Contract № 868/1759-D  dated 3 October 2025). The theoretical investigation of Nikitin-Masalov representation in the current work was supported by the Russian Science Foundation (project number 25-12-00263). The author would like to express his deep gratitude to F.\,Ya.\,Khalili for his invaluable contributions to the research.

	\pagebreak
	\onecolumngrid
	Supplementary Information for ``Protocol for preparing Schr\"odinger-cat states via spontaneous parametric down-conversion and photon number measurement''.
	V. L. Gorshenin\({}^{1,2}\)

	\({}^1\)Russian Quantum Center, Skolkovo 121205, Russia

	\({}^2\)Moscow Institute of Physics and Technology, 141700 Dolgoprudny, Russia

  \newcommand{\plotswidth}{0.47}

  \appendix

  \section{The Nikitin-Masalov representation formulation}

  The Nikitin-Masalov representation \cite{Nikitin_QOpt_JEOS_B_3_2_1991} exploits the conservation of total photon amount wih respect to frequencies \(N\) to transform the parametric-interaction Hamiltonian into a set of Hamiltonians each acting on subspaces with fixed \(N\), with \(N = n_{\rm a} + 2 n_{\rm b}\), of the form

	\begin{equation}
		\hat{H} = \sum_{N=0}^\infty\sum_{k,k'=0}^{[N/2]}
		\ket{N-2k}_s \otimes \ket{k}_p H^{N}_{kk'}\,{}_s\!\bra{N-2k'}\otimes{}_p\!\bra{k'} \,,
	\end{equation}

	where \([\dots]\) denotes the integer part of a number.  The matrices \(H^{N}_{kk'}\) have the following tridiagonal form:
	\begin{equation} \label{tridialgonal-hamiltonian}
		\|\hat{H}^{N}_{kk'}\| =
		\begin{bmatrix}
			0   	& c_0^N	& 0   	&\dots 	& 0 & 0\\
			c_0^N 	& 0		& c_1^N 	&\dots 	& 0 & 0\\
			0   	& c_1^N	& 0   	&\dots 	& 0 & 0\\
			\vdots 	&\vdots &\vdots &\ddots &
			c_{\left[\frac{N}{2}\right]-2}^N & 0\\
			0 		& 0 	& 0 	& c_{\left[\frac{N}{2}\right]-2}^N
			& 0 & c_{\left[\frac{N}{2}\right]-1}^N\\
			0 		& 0 	& 0 	&0 		& c_{\left[\frac{N}{2}\right]-1}^N & 0\\
		\end{bmatrix}
\end{equation}
	with
	\begin{equation}
		c^N_k = \sqrt{(k+1)(N-2k)(N-2k-1)} \,,
	\end{equation}

	Given the initial state of interest vacuum in the signal mode and a coherent state in the pump mode the dynamics could be considered only for fixed \(N\) separately. For each such subspace, the corresponding tridiagonal matrix \eqref{tridialgonal-hamiltonian} has a nondegenerate spectrum that is symmetric about zero. Whether \(\lambda = 0\) is included among the eigenvalues depends on the matrix dimension. This type of spectral symmetry for \eqref{SHG-hamiltonian} was discussed and illustrated in Ref.~\cite{Nikitin_QOpt_JEOS_B_3_2_1991, Alvarez_JPhysA_28_20_1995}.

	\begin{itemize}
		\item
		\(0, \pm |\lambda^{N, j}|, \, j = 0,1,..., (N)/4-1,\,|\lambda^{N, j}|>0\), if \(N/2\) -- even
		\item
		\(\pm |\lambda^{N, j}| \, j = 0,1,..., N/4, \,|\lambda^{N, j}|> 0\), if \(N/2\) -- odd
	\end{itemize}

	The eigenstate denoted by \(\ket{\chi^{N, j}}\) in the Nikitin-Masalov representation with the corresponding eigenvalue \(\lambda^{N, j}\) is the follows:
	\begin{equation}
		\hat{H}^{N} \ket{\chi^{N, j}} = \lambda^{N, j} \ket{\chi^{N, j}}
	\end{equation}

	We further introduce the components of the Hamiltonian eigenstates in the Nikitin-Masalov representation:
	\begin{equation}\label{eq:eigenvector-components}
		\ket{\chi^{N, j}} = \sum_{k=0}^{k=[N/2]} \chi^{N, j}_k \ket{N-2k}_{\rm s}\ket{k}_{\rm p}
	\end{equation}

	The importance of obtaining eigenvalues \(\lambda^{N,j}\) also lies in the ability to obtain eigenstates \(\ket{\chi^{N, j}}\) in the Nikitin-Masalov representation in a simple way using recurrence relations. This is possible due to the tridiagonal form of the Hamiltonian \eqref{tridialgonal-hamiltonian}:

	\begin{equation}
		\chi^{N, j}_{k+1} = \big(\lambda^{N,j} \chi^{N, j}_k - c^{N}_{k-2} \chi^{N, j}_{k-1}\big)/ c^{N}_{k-1}
	\end{equation}
	The eigenstates can be normalized once their components are obtained; thus, we begin by determining the eigenvalues.

	The spectrum of this Hamiltonian has been investigated using semiclassical techniques \cite{Alvarez_JPA_28_20_1995} as well as group-theoretic methods \cite{Mohmadian_PhysLet_A_384_3_2020, Debergh_JPA_MatAndGen_33_40_2000}. In particular, Ref.~\cite{Alvarez_JPA_28_20_1995} provides analytical approximations for the eigenvalues in the large-\(|\lambda|\) regime, i.e., for those with the largest modulus. In Ref.~\cite{Karassiov_PhysLettA_295_5_2002} was considered approximation of all eigenvalues. Unfortunately error of this method was high for eigenvalues near zero excapt straight zero eigenvalue.

	A key advantage of the Nikitin-Masalov representation is that it reduces the eigenproblem to independent fixed-\(N\) subspaces, greatly simplifying the computation of eigenvalues and eigenstates. With a truncation at \(N_{\rm cutoff}\), the effective matrix dimension scales as \(N^2_{\rm cutoff}\) rather than \(N^4_{\rm cutoff}\). The price is that one must diagonalize \(N_{\rm cutoff}\) separate Hamiltonians \(\hat{H}^{N}\); however, each block is much smaller, and the tridiagonal structure makes the numerical eigenvalue and eigenvector computation straightforward.

	\section{Assessment of the validity of truncating the subspace expansion to eigenstates with eigenvalues near zero}

	We introduce a deviation \(\Delta\) to quantify the error incurred when approximating this expansion using only a finite subset of eigenvectors.

	\begin{equation} \label{eq:delta-def}
		\Delta \equiv 1 - \sum_{j=[N/4]-\rho}^{j=[N/4]+\rho} \Big|\bra{\chi^{N, j}} \ket{0}_{\rm s} \otimes \ket{n}_{\rm_p}\Big|^2
	\end{equation}

\begin{figure}
		\centering
		\includegraphics[width=\plotswidth\linewidth]{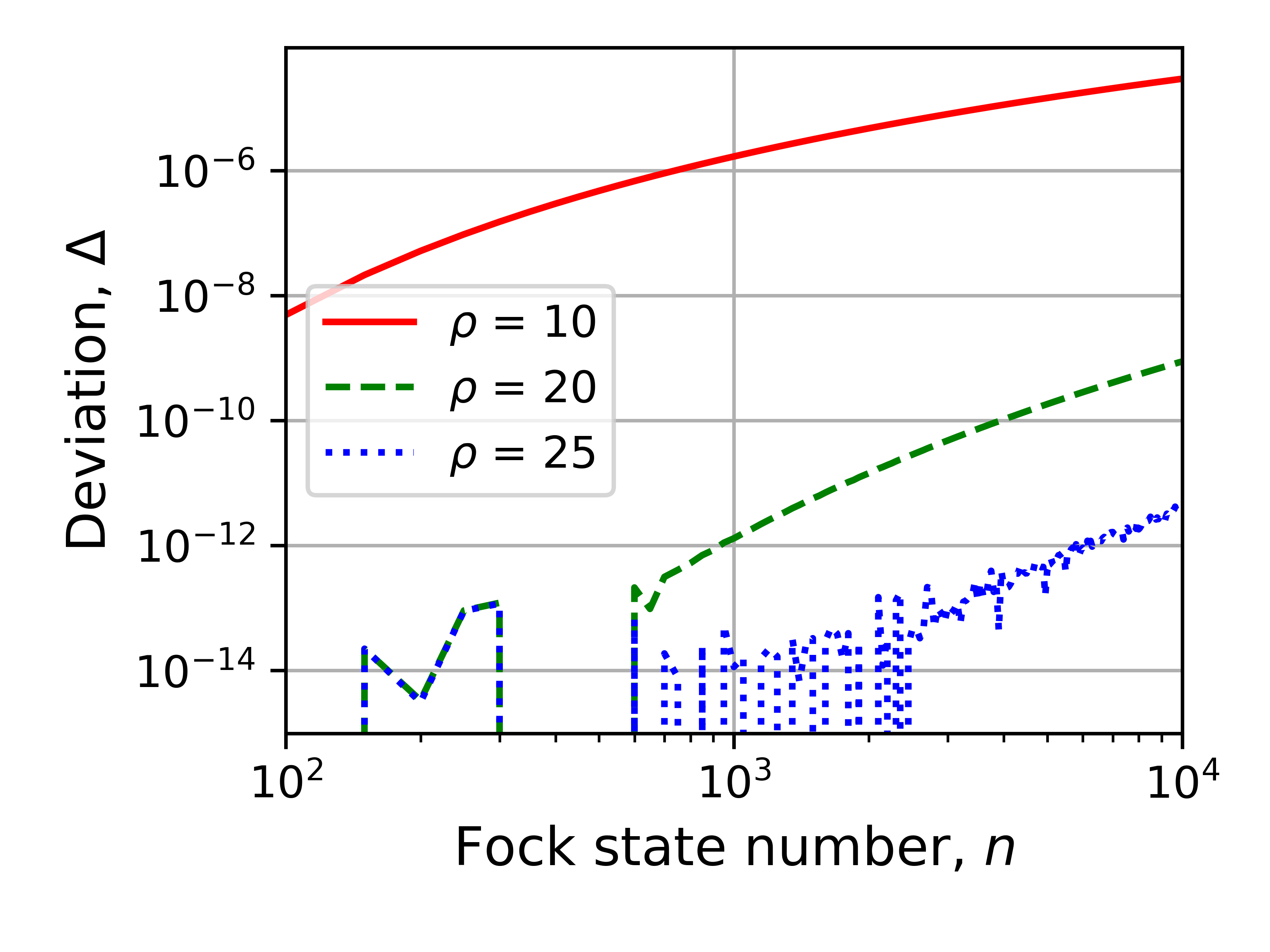}
		\caption{
			Deviation \(\Delta\) incurred when representing the initial state \(\ket{0}_{\rm s} \otimes \ket{n}_{\rm_p}\) using only eigenstates with eigenvalues within a radius \(r\) of zero (see Eq.\,\eqref{eq:delta-def}).
					}
		\label{fig:fock-state-error-vicinity}
	\end{figure}

	If the expansion is performed over the full eigenbasis, the deviation vanishes \(\Delta = 0\). Figure \ref{fig:fock-state-error-vicinity} plots \(\Delta\) as a function of the neighborhood radius \(r\) used to retain eigenvalues near zero. For \(\rho = 25\), deviation \(\Delta\) becomes negligibly small, indicating that the dynamics is accurately captured by a truncated set of eigenvectors. This truncation greatly reduces the computational cost: instead of summing \(2n\) ontributions for the evolution of each two-mode Fock state  \(\ket{0}_{\rm s} \otimes \ket{n}_{\rm p}\) (which correponds \(N = 2n\) in Eq.\,\eqref{eq:delta-def}), we sum at most 51 terms (corresponding to \(\rho = 25\)). In this Letter we therefore fix \(\rho = 25\) throughout. The resulting speedup is crucial in the bright-state regime, e.g., for \(\beta > 30\), where the pump-mode mean photon number \(\langle \hat{n} \rangle = \beta^2\) is close to \(10^3\). The computational cost of evaluating the evolution scales accordingly: for the Fock state \(n\), the direct calculation requires summing \(2n\) terms obtained from the eigenvectors and the corresponding time-evolution phases. In contrast, in the present approach the number of required terms is much smaller, namely \(2 \rho \leq 51 \ll \beta^2 \sim 10^3\).

	\section{Numerical approximation of near-zero eigenvalues in the Nikitin-Masalov representation}

\begin{figure}
		\centering
		\includegraphics[width=\plotswidth\linewidth]{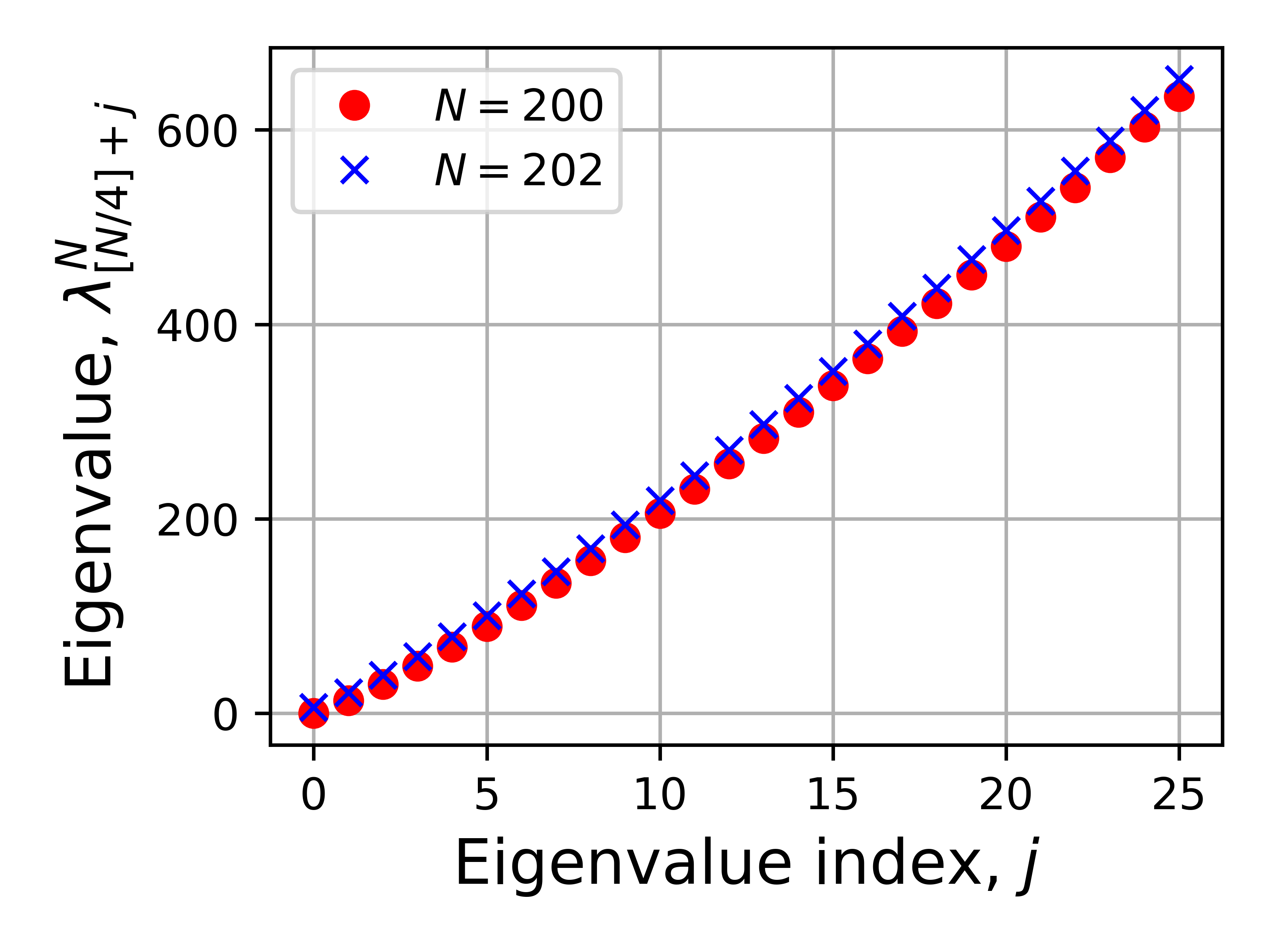}
		\includegraphics[width=\plotswidth\linewidth]{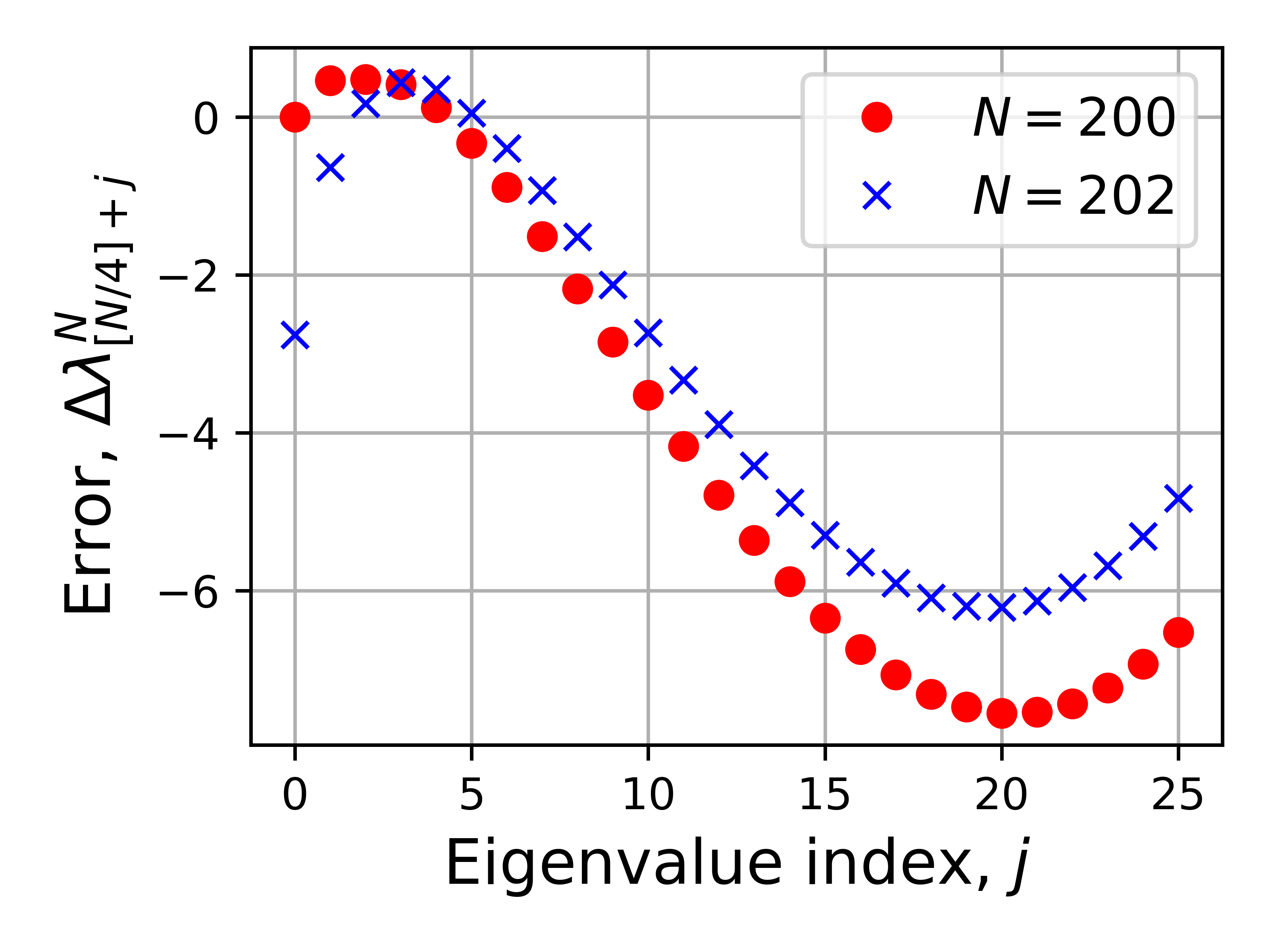}
		\caption{
		(Left) Near-zero eigenvalues as functions of the index for the even series (\(N/2=100\), red circles: exact eigenvalues) and the odd series (\(N/2=101\), blue crosses: exact eigenvalues), with the indexing chosen to start from the zero eigenvalue. (Right) Absolute error of the eigenvalue approximation for the even series (\(N/2=100\), red circles) and the odd series (\(N/2=101\), blue crosses).
		}
		\label{fig:eiv-approximation}
	\end{figure}

	We focus on the eigenvalues in the vicinity of zero. As discussed above, zero is an eigenvalue when \(N/2\) is even, but not when \(N/2\) is odd. The left panel of Fig. \ref{fig:eiv-approximation} illustrates the near-zero behavior for the lowest few eigenvalues. The eigenvalues near zero can be approximated with high accuracy as follows:
	\begin{equation}
		\begin{aligned}
			\lambda^{N,[N/4]+k}\big\vert_{\rm N/2\,-\,even} = b_e(N) k^{d_e(N)}
			,\quad
			\lambda^{N,[N/4]+k}\big\vert_{\rm N/2\,-\,odd} = a_o(N) + b_o(N) k^{d_o(N)}
		\end{aligned}
	\end{equation}

	Here,  \(k = 0,1,..., \rho\), and throughout this section we use \(\rho = 25\) in the approximations. Restricting to the case of even \(N/2\), the coefficients depend on the total energy \(N\)) as follows:

	\begin{equation}
		\begin{aligned}
			b_e(N) \approx 1.2 \cdot (N + 23.5)^{0.43}
			,\quad
			d_e(N) \approx 1.1 + 0.8 \cdot N^{-0.32}
		\end{aligned}
	\end{equation}

	For odd \(N/2\), the dependence of the coefficients on the total energy \(N\) is given by:
	\begin{equation}
		\begin{aligned}
			a_o(N) \approx 3.7 + 0.48 \cdot N^{0.45}
			,\quad
			b_o(N) \approx 1.3 \cdot (N + 14.9)^{0.43}
			,\quad
			d_o(N) \approx -0.34 + 0.89 \cdot N^{-0.34}
		\end{aligned}
	\end{equation}

	For approximation investigation we introduce eigenvalue error \(\Delta \lambda^{N}_{[N/4]+k}\) as follows:
	\begin{equation}
		\Delta \lambda^{N,[N/4]+k} \equiv \lambda^{N,[N/4]+k} \big|_{\rm fine}- \lambda^{N,[N/4]+k} \big|_{\rm approx}
	\end{equation}
	The right panel of Fig. \ref{fig:eiv-approximation} displays the approximation error for \(N=200\) and \(N=202\). The error is clearly much smaller than the spacing to the neighboring eigenvalues. As a result, the approximate eigenvalue and its associated eigenvector provide a suitable starting point for a refinement procedure that reduces the difference between the approximate and exact eigenvalues and yields an accurate eigenvector. Because, the eigenvalues themselves show a more intricate dependence on \(N\), since the deviation in Fig. \ref{fig:eiv-approximation} (right panel) is has complicated dependency.

	\section{Derivation of Eqs\,\eqref{eq:n-avg-cat} and \eqref{eq:dispersion_n_approx}}
	\label{app:medium_and_variance_derivation}

	We calculate the mean photon number \(\langle \hat{n} \rangle\) and the photon-number variance \((\Delta\hat{n})^2\) for the squeezed SC state \(\ket{\psi_{\rm SSC}}^{(0)}\) (see Eq.\,\eqref{eq:cat-state-wavefunction-approx}). Throughout, we adopt the following definitions of the squeezing and displacement operators:

	\begin{equation}
		\hat{S}(r) = \exp \Big(-\frac{r}{2} (\hat{a}^2 - \hat{a}^{\dagger 2})\Big)
		,\quad
		\hat{\mathcal{D}}(\alpha)
		=
		\exp\big(\alpha \hat{a}^\dagger - \alpha^* \hat{a}\big)
	\end{equation}

	together with the following standard property of squeeze and displacement operator:

	\begin{equation}
		\hat{S}(r)^\dagger \hat{a} \hat{S}(r) = \cosh(r) \hat{a} - \sinh(r) \hat{a}^\dagger
		,\quad
		\hat{\mathcal{D}}^\dagger(\alpha) \hat{a} \hat{\mathcal{D}}(\alpha)
		=
		a + \alpha
	\end{equation}

	To evaluate the mean photon number of the SSC state, we first consider the expectation value:

	\begin{equation}
		\begin{aligned}
			V_{\beta,\,\alpha,\,r} =
			\bra{\beta} \hat{S}(r)^\dagger \hat{S}(r) \ket{\alpha}
			=
			\bra{0} \hat{\mathcal{D}}^\dagger(\beta) \hat{S}(r)^\dagger \hat{S}(r) \hat{\mathcal{D}}(\beta) \hat{\mathcal{D}}(\alpha - \beta) \ket{0}
			= \\ =
			\bra{0}
			\Big(e^{-2r} \beta (\beta + \hat{a} + \hat{a}^\dagger)
			+ \cosh(2r) \hat{a}^\dagger \hat{a}
			-  \frac{\sinh(2r)}{2}(\hat{a}^2 + \hat{a}^{\dagger 2}) + \sinh^2(r)
			\Big)
			\ket{\alpha - \beta}
			= \\=
			e^{-(\alpha -\beta )^2} \Big(-\left(\alpha ^2+\beta ^2\right) \frac{\sinh (2r)}{2}
			+
			\alpha \beta  \cosh (2 r)
			+
			\sinh ^2(r)\Big)
		\end{aligned}
	\end{equation}

	We can now use this identity to evaluate the mean photon number of the SSC state as:

	\begin{equation}
		\begin{aligned}
			\langle \hat{n} \rangle
			=
			\frac{
				V_{\alpha,\,\alpha, r} + V_{\alpha,\,-\alpha, r} + V_{-\alpha,\,\alpha, r} +V_{-\alpha,\,-\alpha, r}}{2(1+e^{-2 \alpha^2})}
		\end{aligned}
	\end{equation}

	In the bright limit relevant here (\(\alpha \gtrsim 30\)), terms proportional to \(e^{-2\alpha^2}\) and \(\alpha^2 e^{-2\alpha^2}\) are exponentially suppressed and can be omitted. Equivalently, we neglect the cross terms \(V_{\alpha,\,-\alpha,\,r}\) and \(V_{-\alpha,\,\alpha,\, r}\):

	\begin{equation}
		\begin{aligned}
			\bra{\psi_{\rm SSC}^{(0)}} \hat{n} \ket{\psi_{\rm SSC}^{(0)}} 
			=
			-\alpha ^2 \sinh (2 r)
			+
			\cosh(2r) \alpha^2 + \sinh(2r)
		\end{aligned}
	\end{equation}

	In the case of \(e^{|r|} \ll |\alpha|\), the mean photon number simplifies further to:

	\begin{equation}
		\bra{\psi_{\rm SSC}^{(0)}} \hat{n} \ket{\psi_{\rm SSC}^{(0)}} = e^{-2 r} \alpha^2
	\end{equation}

	Proceeding analogously, we obtain the photon-number variance for \(\ket{\psi_{\rm SSC}^{(0)}}\):

	\begin{equation}
		(\Delta n)^2 \equiv
		\bra{\psi_{\rm SSC}^{(0)}} \hat{n}^2 \ket{\psi_{\rm SSC}^{(0)}}
		-
		\big(\bra{\psi_{\rm SSC}^{(0)}} \hat{n} \ket{\psi_{\rm SSC}^{(0)}}\big)^2
	\end{equation}

	Carrying out the corresponding (lengthy) algebra and taking the large-\(\alpha\) limit (\(\alpha \gtrsim 30\)), we obtain the following accurate approximation:

	\begin{equation}
		(\Delta n)^2  = \alpha^2 e^{-4r}
	\end{equation}

	\section{Probability peaks time approximation by initial coherent state amplitude} \label{app:time-of-peaks-measure-nonzero-photon-number}

	Following the approach of our previous work, we parameterize (i) the peak success probability \(p^{(m)}\) for heralding on outcome \(m\) and (ii) the corresponding optimal interaction time \(\tau_{\rm opt}^{(m)}\) at which this peak occurs via the following empirical expressions:

	\begin{equation}
		p^{(m)} \approx \frac{b_p^{(m)}}{\beta^{d_p^{(m)}}}
	,\quad
		\tau_{\rm opt}^{(m)} \approx \frac{b_\tau^{(m)}}{\beta^{d_\tau^{(m)}}}
	\end{equation}

	The fit coefficients extracted from the numerically computed projection-measurement probabilities are summarized in Table \ref{tab:p-meas-and-tau-opt-approx-coefficients}.

	\begin{table}
	\begin{tabular}{|c|c|c|c|c|c|}
			\hline
			\(m\) &
			$b_\tau^{(m)} $, \% & $d_\tau^{(m)}$ &
			$b_p^{(m)}$ & $d_p^{(m)} $, \% \\
			\hline
			0 & 1.19 & 0.77 & 1.10 & 0.97\\
			\hline
			2 & 1.22 & 0.78 & 0.55 & 0.97\\
			\hline
			4 & 1.26 & 0.79 & 0.41& 0.97\\
			\hline
			6 & 1.31 & 0.80 & 0.35 & 0.98\\
			\hline
		\end{tabular}
	\caption{
		Fit coefficients for the dependence of the success probability \(p^{(m)}\) (heralding on \(m\) detected photons) and the corresponding optimal time \(\tau_{\rm opt}^{(m)}\) on the initial coherent amplitude \(\beta\), for various outcomes \(m\).
				}
	\label{tab:p-meas-and-tau-opt-approx-coefficients}
	\end{table}

	We note that the fit reported previously in Ref.\,\cite{Gorshenin_JOSA_B_42_2_2025} for \(m=0\) differs from the present parameterization. Nevertheless, over the range \(6 \le \beta \le 100\) the two approximations give numerically close values.

	\section{Approximation of prepared Scr\"odinger cat states parameters after measurement small even photon number}
	We extract the SSC state parameters by numerically fitting the two defining quantities of the SSC description: the coherent amplitude \(\alpha^{(m)}\) and the squeezing parameter \(r^{(m)}\). The reported parameter dependences correspond to operating at the interaction time that maximizes the success probability for the measurement outcome \(m\).

\begin{figure}
		\centering
		\includegraphics[width=\plotswidth\linewidth]{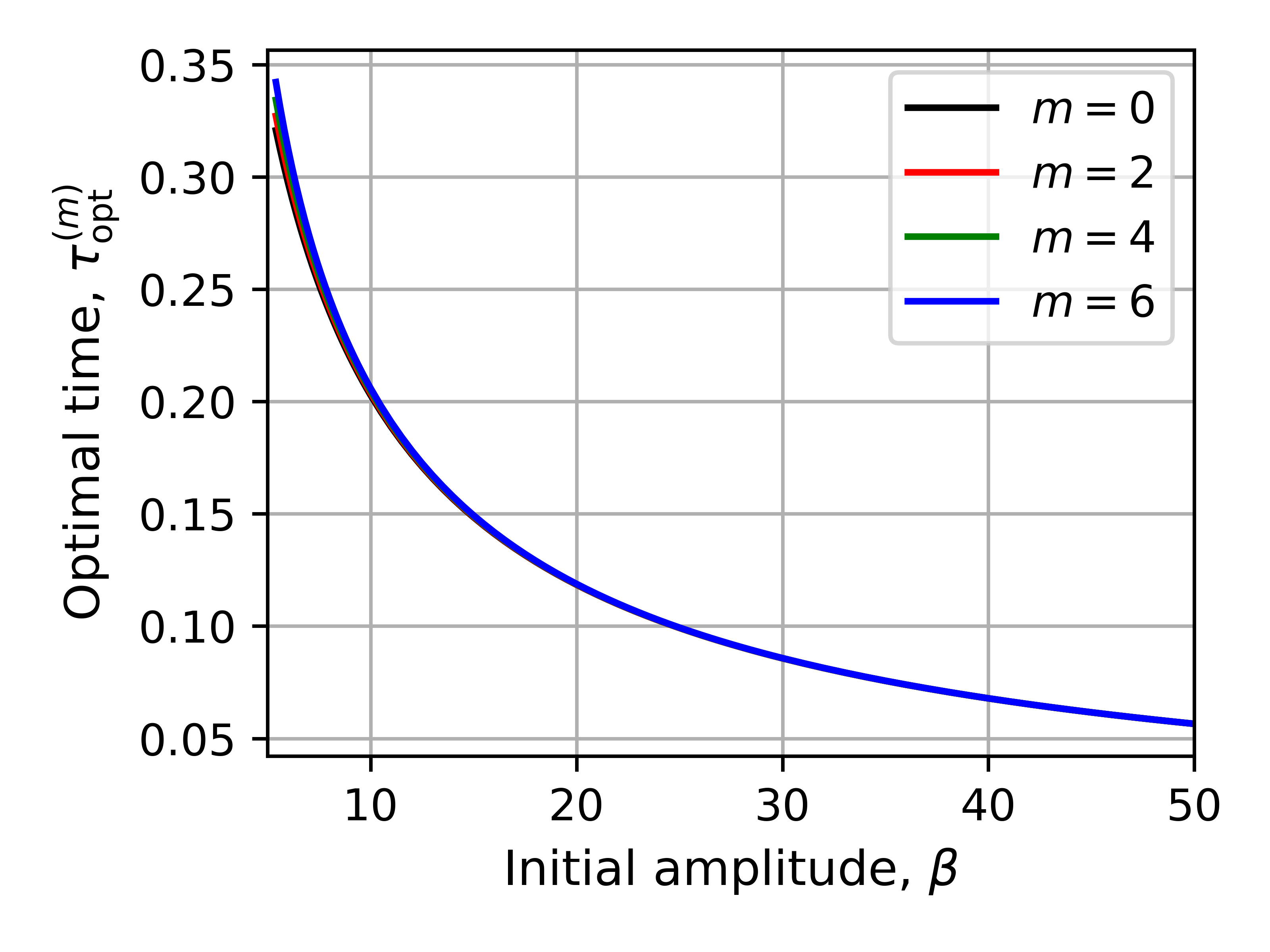}
		\includegraphics[width=\plotswidth\linewidth]{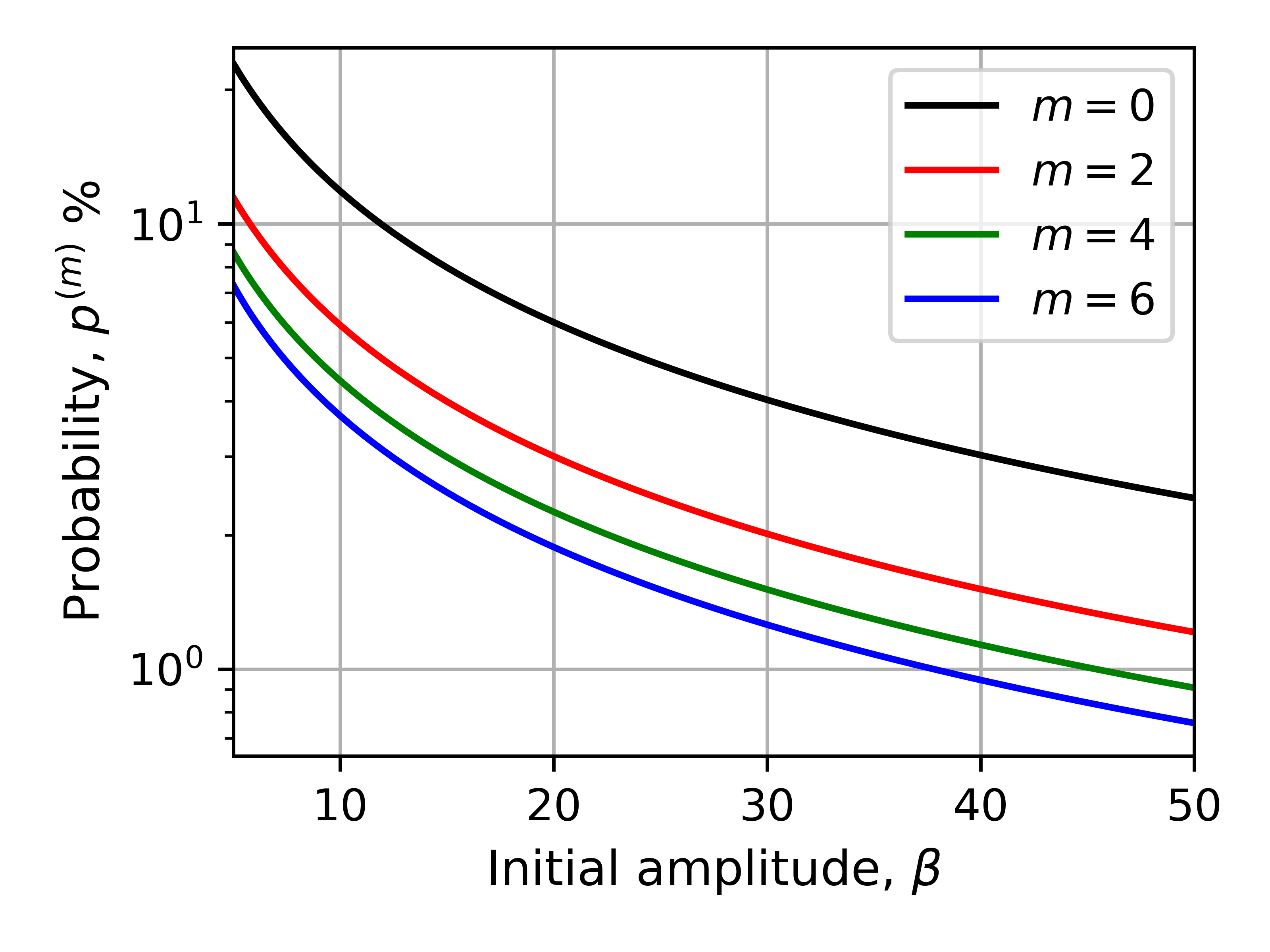}
		\caption{(Left) Optimal time of interaction \(\tau_{\rm opt}^{(m)}\) to  obtain higher possible probability of \(m\) photon measurement by initial coherent state amplitude \(\beta\) in the pump mode. (Right) Corresponding higher achievable probability by initial coherent state amplitude \(\beta\) in the pump mode. Here \(m = 0, 2, 4, 6\).}
		\label{fig:m-meas-time-and-prob-plot}
	\end{figure}

	Figure \ref{fig:m-meas-time-and-prob-plot} summarizes the operating point for heralding on a pump-mode outcome \(m\). The left panel shows the optimal interaction time \(\tau_{\rm opt}^{(m)}\) that maximizes the probability of measuring \(m\) photons, plotted versus the initial coherent amplitude \(\beta\). The dependence of  \(\tau_{\rm opt}^{(m)}\) on \(m\) is weak. The right panel shows the corresponding peak success probability \(p^{(m)}\) evaluated at \(\tau_{\rm opt}^{(m)}\) as a function of \(\beta\). Although \(p^{(m)}\) decreases as \(m\) increases, it remains substantial for experimentally relevant parameters.

\begin{figure}
		\centering
		\includegraphics[width=\plotswidth\linewidth]{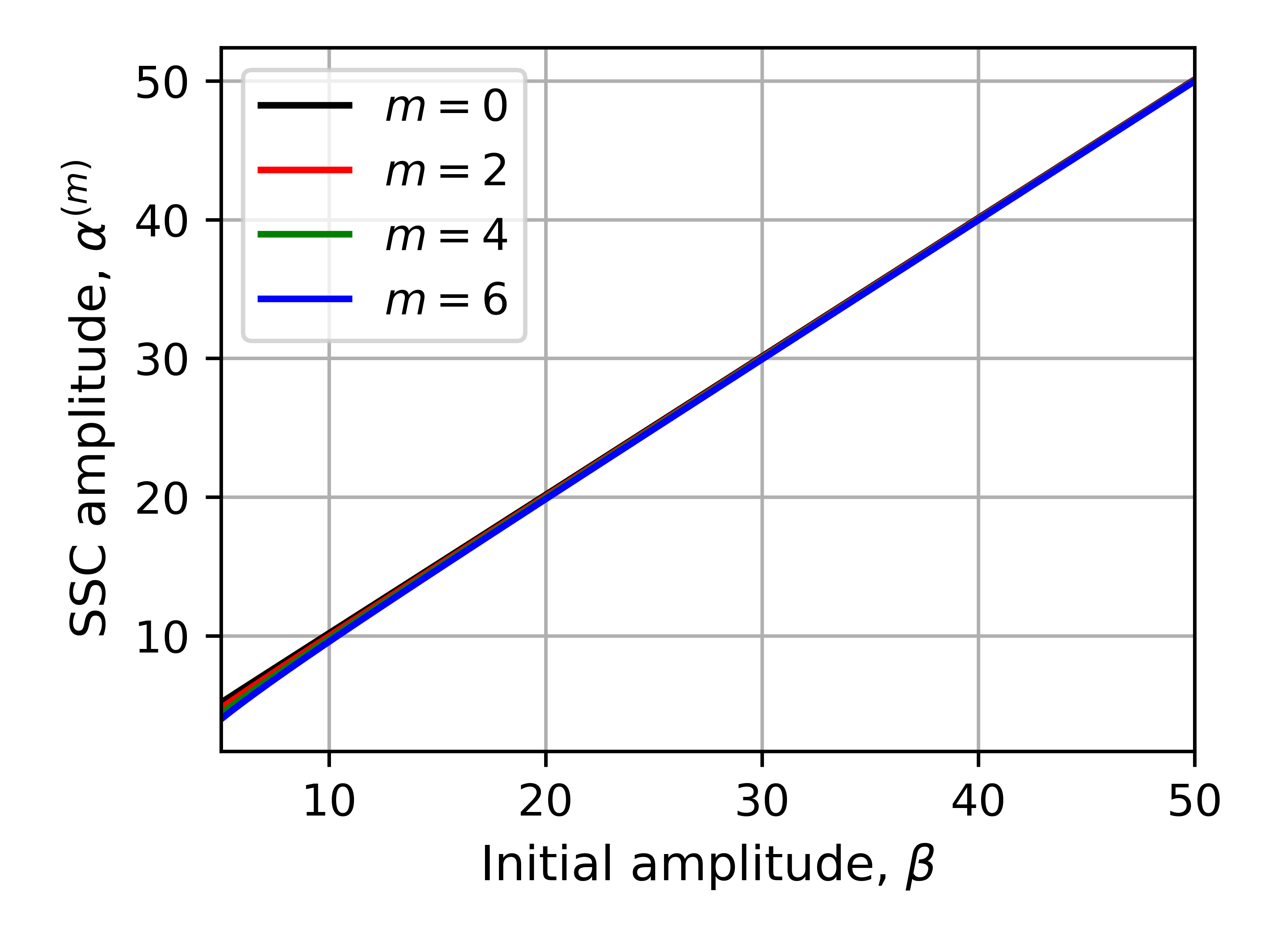}
		\includegraphics[width=\plotswidth\linewidth]{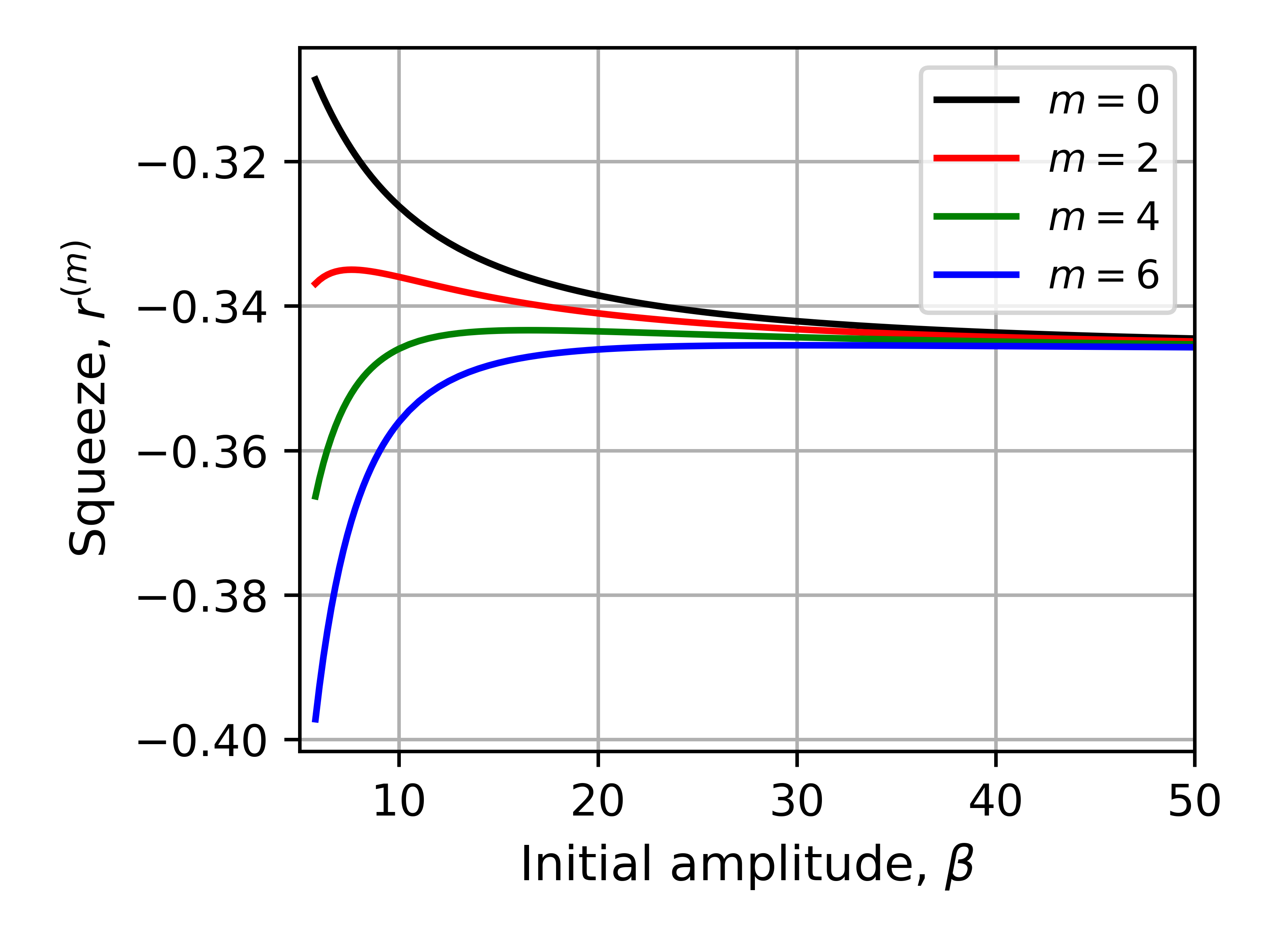}
		\caption{(Left) Amplitude of SSC \(\alpha^{(m)}\) achieving highest possible fidelity after optimal interaction and successful measurement of \(m\) photons by initial coherent state amplitude \(\beta\) in the pump mode. (Left) Squeezing factor of SSC \(r^{(m)}\) achieving highest possible fidelity after optimal interaction and successful measurement of \(m\) photons by initial coherent state amplitude \(\beta\) in the pump mode. Here \(m = 0, 2, 4, 6\).}
		\label{fig:m-meas-alpha-and-sq-plot}
	\end{figure}

	Figure \ref{fig:m-meas-alpha-and-sq-plot} reports the SSC state parameters extracted at the point of maximal fidelity. The left panel shows the fitted coherent amplitude  \(\alpha^{(m)}\) versus the initial pump amplitude \(\beta\). Its dependence on the conditioned photon number \(m\) is weak, and for small \(m\) one finds the simple scaling \(\alpha^{(m)} \approx \beta\). The right panel shows the corresponding squeezing parameter \(r^{(m)}\) as a function of \(\beta\). While the different \(m\) curves converge asymptotically at large \(\beta\), they can differ appreciably at more moderate \(\beta\).

	Because we neglect dissipation and external driving and the dynamics admits a conserved quantity, the total energy is conserved. Accordingly, the total energy of the two-mode system (with the mode-frequency weighting) is the same before and after the interaction. For the initial coherent pump state of amplitude \(\beta\), the initial value is \(N_{\rm init} = 2 \beta^2\). We define the final energy as:

	\begin{equation}
  	E_{\rm final} = \frac{1}{2 (1 + e^{-2 (\alpha^{(m)})^2})} \Big(\bra{\alpha^{(m)}} + \bra{-\alpha^{(m)}}\Big) \hat{S}^\dagger(r^{(m)}) \hat{a}^\dagger \hat{a} \hat{S}(r^{(m)}) \Big(\ket{\alpha^{(m)}} + \ket{-\alpha^{(m)}}\Big) + 2 m
	\end{equation}

	Furthermore, because \(\alpha^{(m)} \approx \beta\) for \(\beta > 5\), the overlap terms proportional to \(\exp(-2(\alpha^{(m)})^2)\) are exponentially suppressed and can be neglected. The final energy is then well approximated by:

	\begin{equation}
  	E_{\rm final} \approx \alpha ^2 e^{-2 r} + 2m
	\end{equation}

 	For an order-of-magnitude estimate, we take \(e^{-2 r^{(m)}} \approx 2\). In the relevant parameter regime, the squeezing contribution \(\sinh(r^{(m)})^2\) is subleading compared with \((\alpha^{(m)})^2\), so we may neglect it to leading order. This yields:

	\begin{equation}
  	E_{\rm final} \approx 2 (\alpha^{(m)})^2 + 2m
	\end{equation}

	This yields a direct relation between the initial pump amplitude \(\beta\) and the SSC state amplitude \(\alpha\):

 	\begin{equation}
  	2 \beta^2 = \left(\alpha^{(m)}\right)^2 + 2m
	\end{equation}
	which can be rewritten in the form of Eq.\,\eqref{eq:connection-init-amplitude-and-ssc-amplitude}. 

	Figure \ref{fig:fid-opt-time-e-n-o} shows the fidelity between the output state \(\ket{\psi_{\rm res}^{(m)}}\) and the SSC state as a function of the initial coherent amplitude \(\beta\). The dependence on the conditioned photon number \(m\) is weak, and fidelities above 99.99\% are achieved for \(\beta>7\).

	\section{Estimation of microring microresonators interaction coefficient}

	For spontaneous degenerate parametric down-conversion, the interaction constant can be estimated following Ref.~\cite{C_Okoth_PRA_99_4_2019} as:

	\begin{equation}
	\begin{aligned}
	g_{\rm SPDC}
	=
	\frac{\chi^{(2)} }{4\sqrt{2}} \sqrt{\frac{\hbar \omega^2_s \omega_p }{\varepsilon_0 V_{\rm eff}}} \frac{1}{n^2_s n_p}
	\end{aligned}
	\end{equation}

	Here \(\chi^{(2)}\) denotes the medium’s second-order nonlinear susceptibility; \(\omega_{\rm s}\) and \(\omega_{\rm p}\) are the angular frequencies of the signal and pump modes; \(\varepsilon_0 = 8.8 \cdot 10^{-12} {\rm F/m}\) is the vacuum permittivity; and \(V_{\rm eff}\) is the effective mode volume. This estimate relies on two simplifying assumptions: (i) the signal and pump modes share the same \(V_{\rm eff}\) and (ii) the group velocity is approximated by the speed of light in the medium.

	We consider a lithium niobate microring resonator with nonlinear susceptibility \(	\chi^{(2)}  = 54 \cdot 10^{-12} m^2/V^2\) \cite{Wang_NPJ_9_1_2023}. The signal and pump wavelengths are taken as 1550 nm and 775 nm, respectively. For lithium niobate at these wavelengths, we use the refractive indices  \(n_s = 2.21\) and \(n_p = 2.26\).

	For the mode volume we use the result was reported in Ref.~\cite{Gao_NewJPhys_23_123027_2021},  \(V = 5.2 \cdot 10^{-15} {\rm m^3}\). With these parameters, the inferred SPDC coupling is \(g_{\rm SPDC} = 2.5\cdot10^6\). Reference \cite{Gao_NewJPhys_23_123027_2021} also reports a quality factor \(Q = 1.23 \cdot 10^8\), corresponding to a photon lifetime  \(t_{\rm rel} \approx 10^{-7}\) s. Combining these values yields a dimensionless interaction time of order \(\tau \leq 0.25\).

	For a resonator implementing degenerate four-wave mixing, the interaction constant can be written as:

	\begin{equation}
		 g_{\rm DFWM} = \frac{3 \hbar \omega_s \sqrt{\omega_r \omega_b} \chi^{(3)}}{16 \varepsilon_0  n_r n_b n_s^2 V_{\rm eff}}
	\end{equation}

	For an order-of-magnitude estimate, we take the three relevant modes to be nearly frequency- and index-matched,  \(\omega_r = \omega_b = \omega_s = \omega\) and \(n_r=n_b=n_s\), which is justified when the free spectral range is small compared with the optical frequency. We assume parameter \(\chi^{(3)} = 4 \cdot 10^{-21} {\rm m^2/V^2}\). Using the geometry reported in Ref.~\cite{Jin_NatPhot_15_5_2021} for a silicon nitride microresonator (\(r \approx 1\,{\rm mm}\), \(h \approx 40\,{\rm nm}\), \(w \approx  5.6\,{\rm um}\)), we estimate the mode volume as \(V \approx 2 \pi r h w = 1.4 \cdot 10^{-15} {\rm m}^3\). With representative material and cavity parameters, this yields a degenerate four-wave-mixing coupling \(g_{\rm DFWM} = 0.6\). The same reference reports \(Q \approx 1 \cdot 10^8\) near 1550 nm, corresponding to a cavity decay rate \(\kappa = 10^7\).

	In second-quantized form, the degenerate four-wave-mixing Hamiltonian is
	\begin{equation}
		\hat{H} = g_{\rm DFWM} \big(\hat{a}^{\dagger 2} \hat{b} \hat{b}_{\rm nd} + \hat{a}^{2} \hat{b}^\dagger \hat{b}^\dagger_{\rm nd}\big)
	\end{equation}

	where \(\hat{b},\, \hat{b}_{\rm nd}\) are the annihilation operators of the depleted and undepleted pump modes, respectively. Within the undepleted-pump approximation, the latter mode is treated classically, and its operator is replaced by a complex amplitude. The Hamiltonian for coherent driving of the undepleted pump mode then reads \cite{Gardiner_PRA_31_6_1985}

	\begin{equation}
		H_{\rm drive} = i \mathcal{E} (\hat{b}^\dagger_{\rm nd} - \hat{b}_{\rm nd})
	\end{equation}
	where \(\mathcal{E}\) denotes the drive amplitude and is given by:
	\begin{equation}
		\mathcal{E} = \sqrt{\frac{P \kappa}{\hbar \omega}}
	\end{equation}
	where \(\omega\) is the drive frequency.

	 If one pump mode is treated as a classical undepleted field, its mean energy is assumed to be much larger than that of the signal mode and the depleted pump mode. Under this hierarchy of scales, the pump-mode dynamics can be estimated by neglecting the nonlinear interaction, leaving coherent driving and cavity loss as the dominant contributions.

	\begin{equation}
		\frac{\partial \hat{b}_{\rm nd}}{\partial t} \approx \mathcal{E} - \kappa \hat{b}_{\rm nd}
	\end{equation}

	Taking the quantum expectation value and using \(\bra{\beta_{\rm nd}} \hat{b}_{\rm nd} \ket{\beta_{\rm nd}} = \beta_{\rm nd}\), we obtain the following estimate for the coherent amplitude of the undepleted pump mode:

	\begin{equation}
		0 = \frac{\partial \beta_{\rm nd}}{\partial t} \approx \mathcal{E} - \kappa \beta_{\rm nd}
	\end{equation}

	State-of-the-art microresonators can be driven with pump powers of order \(10^2\,{\rm mW}\) \cite{Jin_NatPhot_15_5_2021}, which implies \(\beta_{\rm nd} \lesssim 2.8 \cdot 10^5\). This, in turn, yields the following effective coupling coefficient for spontaneous degenerate parametric scattering mediated by degenerate four-wave mixing:

	\begin{equation}
		g_{\rm SPDC, effective} = \beta_{\rm nd}  g_{\rm DFWM}  = 1.7\cdot 10^5
	\end{equation}

	Using the photon lifetime \(t_{\rm rel} = 1\cdot10^{-7}\)s, we estimate a maximal dimensionless interaction time of order \(\tau \leq 0.017\).

\end{document}